\documentclass[twocolumn,preprintnumbers,superscriptaddress,amsmath,amssymb]{revtex4}

\usepackage{amssymb}
\usepackage{graphicx}
\usepackage{dcolumn}
\usepackage{bm}
\usepackage{SIunits}
\usepackage{amsmath}
\usepackage{hyperref}
\usepackage{epstopdf}

\begin{document}


\title{Charge transport through ultrasmall single and double Josephson junctions
coupled to resonant modes of the electromagnetic environment}


\author{Yu.~A.~Pashkin}
\email{pashkin@zp.jp.nec.com}\altaffiliation[on leave from ]{Lebedev Physical Institute, Moscow 119991,
Russia}
 \affiliation{NEC Green Innovation Research Laboratories, Tsukuba, Ibaraki 305-8501, Japan}
\affiliation{RIKEN Advanced Science Institute, Tsukuba, Ibaraki 305-8501, Japan}

\author{H.~Im}
\affiliation{RIKEN Advanced Science Institute, Tsukuba, Ibaraki 305-8501, Japan}
\affiliation{Department of Semiconductor Science, Dongguk University, Phil-Dong, Seoul 100-715, Korea}

\author{J.~Lepp\"akangas}
\affiliation{Institut f\"ur Theoretische Festk\"orperphysik, Karlsruhe Institute of Technology, D-76128
Karlsruhe, Germany}

\author{T.~F.~Li}
\affiliation{RIKEN Advanced Science Institute, Tsukuba, Ibaraki 305-8501, Japan}
\affiliation{Institute of Microelectronics, Tsinghua University, Beijing 100084, China}

\author{O.~Astafiev}
 \affiliation{NEC Green Innovation Research Laboratories, Tsukuba, Ibaraki 305-8501, Japan}
\affiliation{RIKEN Advanced Science Institute, Tsukuba, Ibaraki 305-8501, Japan}

\author{A.~A.~Abdumalikov~Jr.}
\affiliation{RIKEN Advanced Science Institute, Tsukuba, Ibaraki 305-8501, Japan}

\author{E.~Thuneberg}
\affiliation{Department of Physical Sciences, University of Oulu, FI-90014 Oulu, Finland}

\author{J.~S.~Tsai}
 \affiliation{NEC Green Innovation Research Laboratories, Tsukuba, Ibaraki 305-8501, Japan}
\affiliation{RIKEN Advanced Science Institute, Tsukuba, Ibaraki 305-8501, Japan}

\date{\today}

\begin{abstract}

\noindent We have investigated charge transport in ultrasmall superconducting single and double Josephson
junctions coupled to resonant modes of the electromagnetic environment. We observe pronounced current peaks
in the transport characteristics of both types of devices and attribute them to the process involving
simultaneous tunneling of Cooper pairs and photon emission into the resonant modes. The experimental data is
well reproduced with the theoretical models.

\end{abstract}

\maketitle

\begin{figure}[t]
\centering
\includegraphics[width=0.42\textwidth]{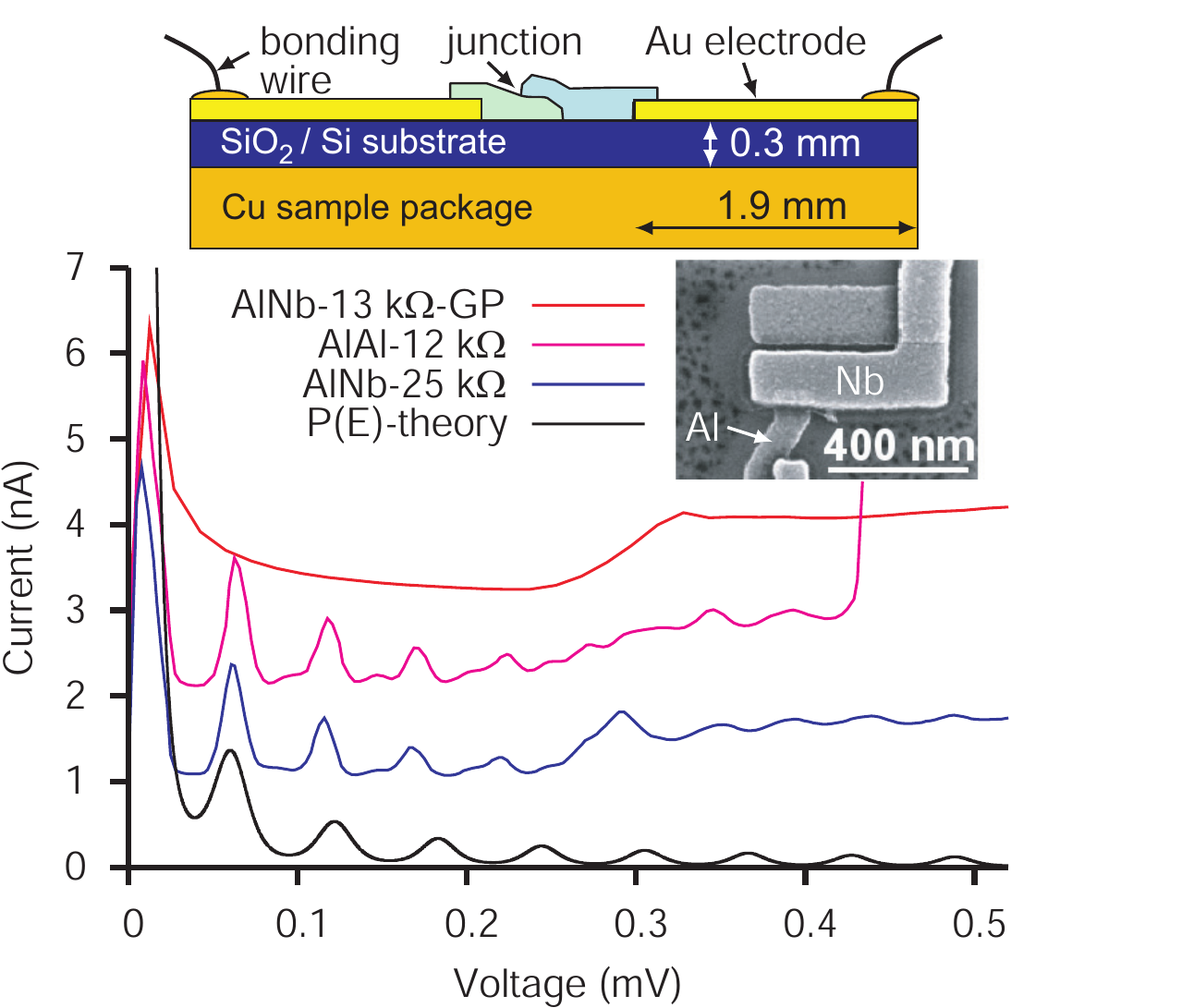}
\caption{(Color online) Current as a function of bias voltage through single Josephson junctions embedded in
a resonant environment and the $P(E)$-theory result with the quasiparticle tunneling neglected. For one
junction, AlNb-13 k$\Omega$-GP, the environmental resonances were suppressed by using a ground plane right
beneath the Au leads. The experimental curves are shifted by 1, 2 and 3~nA, respectively, from bottom to top.
The normal state resistances in the legend were used to deduce the Josephson coupling energies using the
Ambegaokar-Baratoff relation~\cite{AB}, $\Delta_{\rm Al}=215$~$\mu$eV, $\Delta_{\rm Nb}=1.15$~meV and
$T=100$~mK. The Josephson coupling energy of the AlNb-25~k$\Omega$ junction is practically the same as of the
AlAl-12~k$\Omega$ junction ($E_J=57$~$\mu$eV), for which the simulation is made. The EE is characterized in
the text. Inset: chip layout (not to scale) and scanning electron micrograph of the AlNb single junction.}
\label{singledata}
\end{figure}

\indent Charge tunneling in ultrasmall junctions is affected by the electromagnetic environment
(EE)~\cite{IngoldNazarov,SchoenZaikin}.  For example, the same tunnel junction placed in different EEs
exhibits different transport characteristics~\cite{DelsingPRL89}. It was understood that the tunneling charge
probes the electromagnetic environment at certain distances, thus forming a coupled junction+environment
system~\cite{NazarovZETP89}. Similar physics governs the single-electron transistor
(SET)~\cite{IngoldNazarov}, but now resulting in characteristics that are periodically modulated by the gate
voltage.
\newline
\indent For single junctions these ideas were developed into a relatively simple model, nicknamed
$P(E)$-theory ~\cite{DevoretPRL90}, in which the function $P(E)$ describes the probability for the tunneling
electrons (charges) to exchange energy $E$ with the environment. Due to the progress in nanofabrication, it
became possible to engineer on-chip EE by using various techniques. For example, high-impedance environment
can be created by placing miniature thin-film resistors in the dc leads in the vicinity of the tunnel
junction~\cite{HavilandEPL91,KuzminHavilandPRL91}. It is also possible to construct an environment with
distinct, well-characterized resonance modes~\cite{HolstPRL94,Basset}.
\newline
\indent Although the effect of the resonant EE is well understood and received sufficient experimental
support \cite{HolstPRL94,Basset} and theoretical explanations \cite{IngoldGrabert94} in the case of a single
Josephson junction (JJ), clear experimental data for transport properties is still lacking in the case of a
superconducting SET (SSET). The SSET coupled to a lossy transmission line was studied experimentally in
Ref.~\cite{LuMarRimPRB2002}, and some signatures of the environmental modes were observed. In the present
work, we report our transport measurements on single and double Josephson junctions coupled to distinct
electromagnetic modes of the environment. Both types of devices reveal characteristic peak structure in
transport that is interpreted to arise from the interaction of the tunneling Cooper pairs with the resonant
environment. The experimental data of the SSET is well reproduced by the theory applying density matrix to
Cooper pair tunneling in the presence of the (i) environmental modes, identified in the single JJ experiment,
and (ii) substantial subgap electronic density of states in the superconducting Nb island. The observed
resonances appear to be generic for the structures having metallic leads on the surface of, e.g., Si chips,
unless a special care is taken to suppress them.
\newline
\indent The configuration of our chips is shown in the inset of Fig.~1. The samples consist of
Al/AlO$_{x}$/Nb tunnel junctions and Al/AlO$_{x}$/Nb/AlO$_{x}$/Al single-electron transistors made by
electron-beam lithography~\cite{ImAPL06} and embedded into Au leads prefabricated photolithographically on a
silicon chip. In addition to the Al/Nb junctions, an all-Al junction was also measured. The Si chip of
thickness $0.3$~$\milli\meter$ with a 300~nm thick oxidized layer on top is placed in a copper sample package
forming a ground plane at the bottom of the chip. The Au leads and the ground plane form microstrips on both
sides of the junctions/transistors with a wave impedance $Z_0\approx 40$~$\Omega$. The microstrips are bonded
with $0.025$~$\milli\meter$ Au wire to the cryostat wires for dc measurements. The impedance of the bonding
wire ($Z_L\backsim 200$~$\Omega $) creates an impedance mismatch. Therefore, the microstrips resonate
approximately at frequencies $n\omega _{0}$, where $n$ is an integer and $\omega _{0}=\pi
c/\sqrt{\varepsilon_e }\ell $ is the fundamental frequency corresponding to the $\lambda /2$ resonance. Here
$c $ is the speed of light in vacuum, $\varepsilon_e \approx 7$ is the effective dielectric constant of the
microstrip~\cite{Pozar} and $\ell \approx 1.9$~$\milli\meter$ is the microstrip length. These numbers give
the resonance frequency $\omega _{0}/2\pi \approx 30$~$\giga\hertz$. With the resistance of each Au lead
measured at subKelvin temperature $R_{\rm Au}\thickapprox 3.5$~$\Omega$ we estimate the microstrip resonator
quality factor $Q\approx 5$, which is mostly determined by the external impedance of 200~$\Omega$.
Measurements were done using a two-probe scheme in a dilution refrigerator at a base temperature of about
$40$~mK.
\newline
\indent A collection of the single junction dc current-voltage ($I$--$V$) characteristics at the subgap
voltage is shown in Fig.~\ref{singledata}. The common feature of the junctions measured, regardless of the
electrode materials, is that at low bias voltage they all exhibit approximately equidistant peak structure.
The peak position does not depend on the junction resistance. The location of the first peak at a bias
voltage of $\approx 61$~$\mu$V agrees with the estimated resonance frequency $\omega _{0}$\ through the
relation $2eV=\hbar \omega _{0}$ (leading to $\omega_0/2\pi=29.4$~$\giga\hertz$).
\newline
\indent The peaks are suppressed  if an additional ground plane is placed beneath the leads separated by a
thin insulating layer, see Fig.~\ref{singledata}. In this case the Au leads do not form microstrip resonators
as their quality factor drops below one. We also observe a step-like enhancement of the current roughly above
$V=\Delta_{\rm Al}/e$ in both Al/Nb and all-Al JJs, resulting probably from Andreev tunneling~\cite{HekNaz}.
\newline
\indent We model the single JJ $I$--$V$ characteristics using the $P(E)$-theory. The electromagnetic
environment seen by the JJ is a parallel connection of the junction capacitance $C_J$ and transmission line
impedance~\cite{Pozar} $2Z_0[Z_L+Z_0\tanh(\gamma \ell)]/[Z_0+Z_L\tanh(\gamma \ell) ]$, where the factor 2
emerges from the fact that we have effectively two equivalent resonators in series, and $\gamma \ell=R_{\rm
Au}/2Z_0+i\pi \omega/\omega_0$. In the simulations, we have  used the junction capacitance $C_J=1$~fF and the
fundamental resonance frequency $\omega_0/2\pi=29.4$~$\giga\hertz$ assuming also that the resonances are
equidistant. In Fig.~1 we see a reasonably good agreement between the simulated curve and the measured ones
for both peak positions and magnitude. The shift of the higher resonant peaks towards the origin visible in
the experimental curves can be explained by the fact that for the given leads geometry the resonances occur
not at multiples of $\omega_0$ but at lower frequencies. This is confirmed by more detailed microstrip
analysis performed with Microwave Office. Multiphoton transitions (to the fundamental mode) could also result
in a similar peak structure, but are present with negligible magnitudes.
\begin{figure}[t]
\centering
\includegraphics[width=0.33\textwidth,angle=-90]{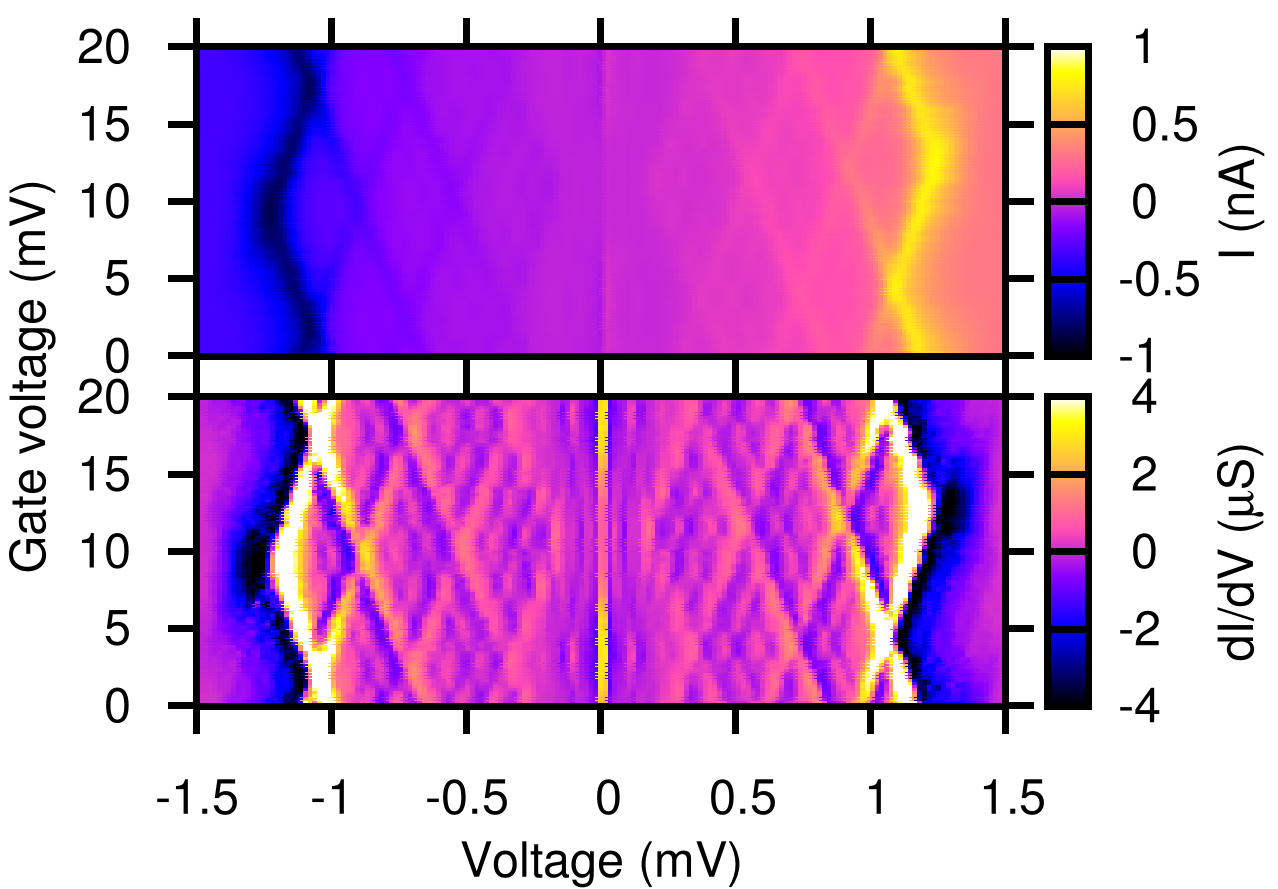}
\caption{(Color online) Current (upper panel) and differential conductance $dI/dV$ (lower panel) through the
Al-Nb-Al SSET as a function of the gate and bias voltages. The characteristics are $e$- periodic as a
function of the gate and consist of rhombi-shape cells, which are due to the subgap JQP cycles (see the
text), ending with intense normal JQP cycles at higher bias voltages. Differential conductance reveals
resonant peak structure in each cell.} \label{largescale}
\end{figure}
\newline
\indent We then turn to characterization of the SSET transport shown in a large scale in
Fig.~\ref{largescale}. The Josephson-quasiparticle (JQP) current~\cite{FultonJQP} is seen at about $V=\pm
1.2$~mV but also rhombi-shape cells exist at lower voltages, starting already from $0.18$~mV, with the gate
modulation period of about 15~mV. The lower voltage structure occurs also due to the JQP processes. Such a
multiple JQP peak structure in Al/Nb SSETs has been observed and discussed in earlier
works~\cite{Harada,Dolata,Toppari}. The new feature here is the structure inside the cells, see also
Fig.~\ref{comparison}a.
\begin{figure}[t]
\centering
\includegraphics[width=0.48\textwidth]{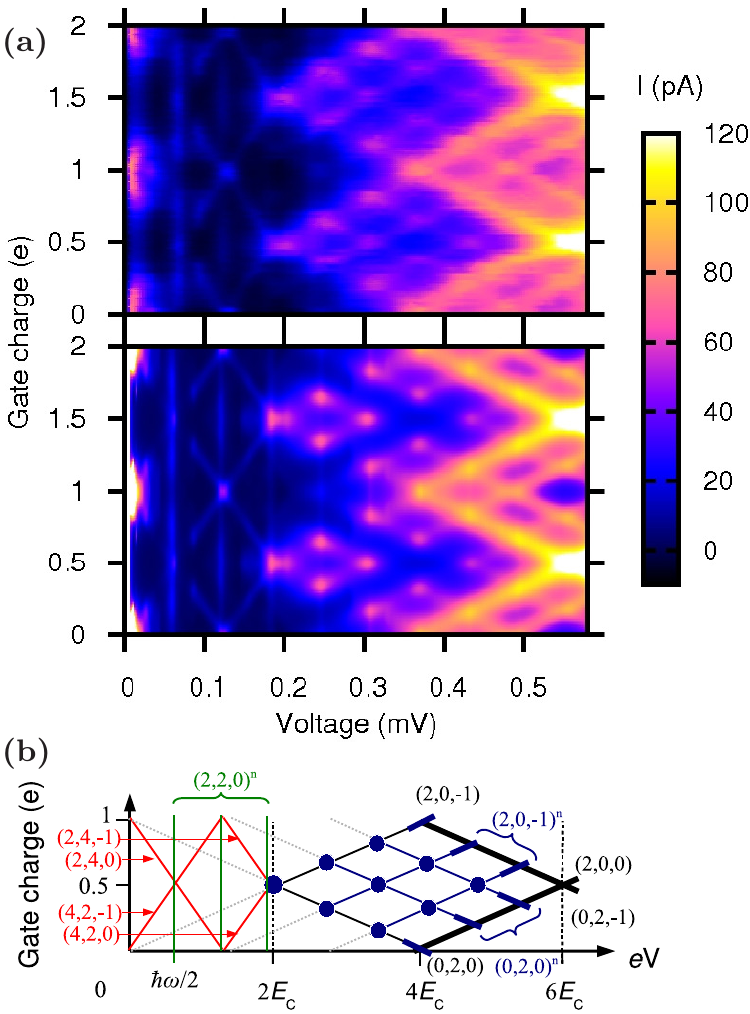} \caption{(Color online) (a): Measured current (upper
panel) shown in Fig.~\ref{largescale} but in a smaller scale, and the current calculated by the DMA (lower
panel). Since the DMA does not hold in the supercurrent region the numerical results are shown from
$10$~$\mu$V. The parameters in the simulation are $\Delta_{\rm Al}=215$~$\mu$eV, $\Delta_{\rm Nb}=1.15$~meV,
 $\Gamma=\Delta_{\rm Nb}/80$, $r=1/45~\mu$s, $T=100$~mK. The EE seen by the SSET is characterized in the text.
(b): Origin of the resonance structure. In a resonant tunneling $(a,b,i)^n$ the charge $ae$ ($be$) tunnels
across the left (right) JJ with the simultaneous excitation of the $n$-th mode, the island charge being
initially $ie$~\cite{LepThunPRB2008}. At low voltages only the third-order CP tunneling resonances are seen
[$(a+b)/2=3$], in addition to cotunneling ($a=b$) with the simultaneous excitation of a corresponding mode.
At higher voltages the subgap JQP cycles, enabled by niobium's subgap DOS, become first possible at double
resonance points (dots) and eventually nearby thick solid lines. }\label{comparison}
\end{figure}
\newline
\indent Our simulations of the SSETs are based on the density-matrix approach (DMA) developed in
Ref~\cite{LepThunPRB2008}, with supplements given below. The model describes charge transport via incoherent
Cooper-pair tunneling taking place at voltages above the supercurrent branch. To account for the subgap
density of states (DOS) in the niobium (island) we use a broadened BCS DOS of the form~\cite{Dynes}
\begin{gather}
n(E)=\left\vert {\rm Re}\left\{\frac{E-i\Gamma}{\sqrt{(E-i\Gamma)^2-\Delta_{\rm Nb}^2}}\right\}\right\vert,
\label{density}
\end{gather}
where $\Gamma$ is a broadening parameter. At low energies such a DOS is a constant, $\Gamma/\Delta_{\rm Nb}$,
resembling the metallic DOS. Further, we assume that an unpaired electron tunneled into the niobium island
relaxes quickly to the Fermi level due to the significant subgap DOS. This kills the asymmetry of the
tunneling rates for different parities of the island~\cite{SchonSiewert} and results in $e$-periodic
characteristics.
\newline
\indent We also see the effects due to non-equilibrium quasiparticles~\cite{MartinisQP} entering the island
from the leads at energy $\Delta_{\rm Al}$ or higher. For qualitative modeling we assume that the electron
tunneling rates into and out of the island are equal and described by the expression~\cite{SchonParity}
\begin{gather}
R(E)=\frac{r}{1+\exp[-(E+\Delta_{\rm Al}-\Delta_i)/k_BT]}, \label{leakage}
\end{gather}
where $\Delta_i$ is the energy gap in the island, $E$ the energy change in the SSET and the maximum rate $r$
characterizes the magnitude of the leakage. In the ideal case practically none of the quasiparticles get
through due to large $\Delta_i=\Delta_{\rm Nb}$, however, due to significant subgap DOS we put $\Delta_i=0$
and calculate the non-equilibrium part of the quasiparticle tunneling using $R(E)$ as a prefactor of the
corresponding Lindblad operators~\cite{LepThunPRB2008}.

Numerical results obtained by the DMA are compared with the experiment in Fig.~\ref{comparison}. The
Josephson coupling energies are deduced from the transport measurements giving $R_1+R_2=73$~k$\Omega$ and
assuming that $R_1=R_2$. The charging energy of the island is obtained from the cell structure giving
$E_C\approx 93$~$\mu$$e$V ($C_1\approx C_2$). As the effective EE seen by the SSET we use practically the
same impedance that fitted the single JJ data but using $C_J=C_1/2$. In the DMA, the environmental impedance
is included directly in the tunneling rate, similarly as in~\cite{HolstPRL94}. This approximation neglects
the rare multiphoton transitions. In order to model (non-Markovian) low-frequency  background charge
fluctuations, we have also averaged the numerical results with respect to the gate charge with a Gaussian
distribution of width $4e/100$.

Let us analyze first the novel features observed in this work. Due to the broadened DOS in the niobium the
onset of the (subgap) JQP cycle occurs when enough energy can be released to create an excitation on the
aluminum side of the JJs. This leads to the threshold $eV=\Delta_{\rm Al}-E_C$ for the double JQP cycle and
$eV=\Delta_{\rm Al}+E_C$ for the ordinary (but subgap) JQP cycle. The resonances appear first at double JQP
points (in Fig.~\ref{comparison} at the low voltage corners of the JQP cells, i.e., at $eV=2E_C\approx
185$~$\mu$eV and $Q_0=e/2+me$) and approximately above the threshold $eV=\Delta_{\rm Al}+E_C$ as continuous
resonance lines. Inside each of the JQP cells there is a fine structure due to the coupling to the resonant
modes of the transmission lines. The origin is that the Cooper-pair tunneling in the JQP process can also be
resonant if the extra energy released is absorbed by the environmental mode. This leads to extra JQP-$n$
lines located at $n\hbar\omega_0/e$  above the main resonant voltages, each corresponding to a photon
emission to the $n$th mode. These extra resonances show similar behavior as the main JQP resonances and
appear first as dots at double resonance points, and as resonance lines at higher bias voltages. The exact
thresholds are $eV=\Delta_{\rm Al}\pm E_C+n\hbar\omega_0/2$ corresponding to double resonances ($-$ sign) and
resonance lines (+ sign). For the double resonance to occur, two simultaneous resonances corresponding to two
different parities of the island charge must coexist, and the resonant states must match. It follows from
this that the double JQP-n resonances can occur only inside the triangle-shaped areas starting at $eV=2E_C$.
The subgap JQP structure disappears finally after the onset of the ordinary JQP cycle at $eV\approx
\Delta_{\rm Al}+\Delta_{\rm Nb}$ (see Fig.~\ref{largescale}). The position of this ordinary JQP is reduced
from the ideal one $\Delta_{\rm Al}+\Delta_{\rm Nb}+E_C<eV< \Delta_{\rm Al}+\Delta_{\rm Nb}+3E_C$ due to the
high subgap leakage of Nb based junctions, which is described by the broadened DOS, and due to quantum Zeno
effect in the charge transport~\cite{LepThunPRB2008}, as nearby the threshold the quasiparticle rate is high
compared to the Josephson coupling energies.

The region of the intermediate voltages (50--200 $\mu$V) reveals resonances due to the Cooper-pair
cotunneling across both JJs and a simultaneous photon emission to the first mode, which enhances the current
at $V\approx 61$~$\mu$V. While the position of the resonance is gate independent, its magnitude increases
when also first- or third-order Cooper-pair tunneling (without photon emission) becomes resonant, see also
Fig.~3b. Also traces of further modes via such cotunneling can be found at multiples of this voltage. The
third-order Cooper-pair tunneling resonances~\cite{LepThunPRB2008}, corresponding to $4e$ tunneling across
one and $2e$ tunneling across the other JJ, are seen as sloping resonance lines in the range $80-180$~$\mu$V.
These resonances were reported also in earlier works~\cite{Billangeon,Joyez}.

Finally, we come to the supercurrent region (0--50 $\mu V$). The current peaks are $e$-periodic, resulting
from the presence of significant subgap DOS. The number of electrons on the island is described by an integer
$m$.  As there is no extra energy related to an odd $m$, the system tends to minimize the electrostatic
energy $(me-Q_0)^2/2C$ through slow higher-order processes. It therefore avoids the degeneracy points $\vert
Q_0-me\vert=e$, which enable large supercurrents across the system. As a result, in the absence of external
quasiparticle leakage, there should be no significant supercurrent, and it has maxima at $Q_0=e/2+me$ (with
integer $m$). This is not the case in the experiment. Indeed, the smallness of the subgap
normal-superconductor tunneling current makes it possible that a small leakage current ($1/r$ between micro-
and millisecond) opposes the slow higher-order processes and provides strong supercurrent peaks at
$Q_0=e+me$.
\newline
\indent In summary, we have observed pronounced current peaks in the SET transport resulting from the Cooper
pair tunneling and simultaneous single photon emission into the resonant modes. They are seen in a wide range
of bias voltages due to the finite electronic subgap DOS in the niobium island. These gate dependent peaks
have the same origin as those observed at a lower bias voltage that do not depend on the gate voltage. Also,
all these peaks have their analogs in transport through single Josephson junctions, which were reported
earlier. Apparently, the observed resonances may be detrimental to various practical devices as they, e.g.,
cause extra decoherence of quantum bits and errors in charge pumps.
\newline
\indent We thank S. Ashhab, D. V. Averin, A. Maassen van den Brink, M. Marthaler, G. Sch\"on, and A. Zagoskin
for fruitful discussions. This work was supported by the JSPS through its FIRST Program, JST-CREST and MEXT
kakenhi ``Quantum Cybernetics".

\end{document}